\documentclass[aps,superscriptaddress,twocolumn,twoside,floatfix,pra,longbibliography,a4paper,nofootinbib,accepted=2023-01-13]{quantumarticle}
\pdfoutput=1
\usepackage{amsmath}
\usepackage{amscd}
\usepackage{wasysym}
\usepackage{color}
\usepackage{braket}
\usepackage{float}
\usepackage{lipsum}
\usepackage[numbers,sort&compress]{natbib}

\usepackage[T1]{fontenc}
\usepackage{ae,aecompl}
\usepackage{amsfonts}
\usepackage{amsthm}
\usepackage{dsfont}
\usepackage{graphicx}
\usepackage{amsfonts}
\usepackage{physics}
\newcommand{\ue}{\mathrm{e}}

\newcommand{\la}{\langle}
\newcommand{\ra}{\rangle}

\newcommand{\R}{\mathbb{R}}

\newcommand{\N}{\mathbb{N}}

\begin{document}

\title{Persistent nonlocality in an ultracold-atom environment}

\author{Bradley Longstaff}
\author{Jonatan Bohr Brask}
\address{Department of Physics, Technical University of Denmark, Fysikvej, 2800 Kgs. Lyngby, Denmark}

\begin{abstract}
We investigate nonlocal quantum correlations arising between multiple two-level impurity atoms coupled to an ultracold bosonic gas. We find that the environment-induced dynamics of the impurity subsystem can generate nonlocal states that are robust against noise and violate a multipartite Bell inequality when projective spin measurements are made. Genuine multipartite nonlocality is also observed in a system of three impurities. We show that non-Markovian effects, and the persistence of coherences in the impurity subsystem, are crucial for preventing complete loss of nonlocality and allow for nonlocal correlations to be generated and maintained for extended periods of time. 
\end{abstract}

\maketitle

Quantum physics allows for correlations that have no classical counterpart. In modern quantum information science, entanglement plays a key role as a resource \cite{Horodecki2009}, enabling communication protocols such as quantum teleportation, quantum-over-classical speed-ups in computation \cite{Jozsa2003}, and measurement precision beyond the classical shot-noise limit \cite{Giovannetti2006}. The strongest quantum correlations defy explanation in terms of a local theory \cite{Einstein1935,Bell1964}. As shown by Bell \cite{Bell1964}, the assumption of local causality places constraints on the possible observations for parties that do not communicate, known as Bell inequalities. These constraints can be violated in quantum mechanics when the parties share an entangled state on which they each make local measurements, thus allowing the observation of Bell nonlocality \cite{Brunner2014}. Nonlocality has been decisively demonstrated in experiment  \cite{Hens15,Shal15,Gius15} and is recognised to enable information processing at an unprecedented level of security, such as device-independent random number generation \cite{colbeckPhD2009,Pironio2010} and quantum key distribution \cite{Pironio09}.

Understanding how and what quantum correlations arise in interacting systems thus presents an intriguing question, and generating and stabilising them has many potential applications. In particular, the study of correlations is key to the characterisation of many-body systems and many-body systems can support a wide range of correlations \cite{deChiara2018}. The problem of generating and detecting nonlocality is related but distinct from that of entanglement, as entanglement is necessary but not sufficient for nonlocality \cite{Werner89,Barrett02}. Nonlocality in many-body systems is more challenging to observe \cite{Tura2014,Zizhu2017,Deng2018,Oudot2019} but does arise, as shown in a number of works, e.g.~\cite{Wasak2018,Fadel2018,Piga2019,Watts2021,Frerot2021,Kitzinger2021,Vieira2021}. At the same time, engineered environments have enabled dissipative entangled state preparation, where the effects of noise and loss are used to an advantage rather than being a hindrance \cite{Kraus08,Verstraete09}. Steady-state entanglement generation is also possible with these techniques \cite{Vacanti09,Aron14,Schneider02,Yuan20} and can even be achieved via incoherent interactions with thermal environments \cite{Plenio02,Brask15}. In the latter case, it has recently been suggested that the generated entanglement be quantified based on its usefulness for particular tasks, including Bell inequality violation \cite{Brask21}. However, similar techniques for nonlocality generation remain largely unexplored \cite{Zou21}.

In this work we consider the generation of nonlocality between two-level impurity atoms coupled to an ultracold bosonic gas, which is trapped in a one-dimensional optical lattice and described by the Bose-Hubbard model \cite{Jaksch98}. The subsystem of impurity atoms can be viewed as an open system in a controllable environment \cite{Cosco18}. Previous analyses of a single impurity atom embedded in a Bose-Hubbard lattice have found that the pure-dephasing dynamics of the impurity can be non-Markovian \cite{Cosco18,Caleffi21,Haikka11}. This observation partially motivates our choice of this model for nonlocality generation. Suppose that the system-environment interactions can generate nonlocality in the impurity subsystem. By viewing non-Markovianity in terms of information backflow from the environment to the system \cite{Breuer09}, one might suspect that memory effects could suppress the destruction of nonlocal correlations due to decoherence. Indeed, previous work examining entanglement generation between two impurities reported on the possibility of entanglement trapping \cite{McEndoo13}, where non-Markovian effects completely halt the dephasing and thus the loss of entanglement. 

Here we demonstrate that memory effects can lead to persistent nonlocality, in which the reduced state of the impurity atoms evolves from a separable state into a nonlocal state and remains nonlocal for an extended period of time. Not only is the resulting nonlocality found to be robust against noise but it also survives in large systems. Note that the focus is on nonlocality generation without the precise dynamical control usually involved in realising gates \cite{Cirac95,Molmer99}. In this respect, our work is similar in spirit to autonomous steady-state entanglement generation \cite{Vacanti09,Aron14,Schneider02,Yuan20,Plenio02,Brask15}.

A system of $N$ ultracold atoms in a one-dimensional optical lattice is well described by the Bose-Hubbard Hamiltonian 
\begin{equation}
	\label{eqn:BH Hamiltonian}
	\hat{H}_{BH} = -J\sum_{j=1}^{M}\left(\hat{a}_{j+1}^\dag \hat{a}_j + \hat{a}_j^\dag \hat{a}_{j+1}\right) + \frac{U}{2}\sum_{j=1}^{M} {\hat{a}_j^\dag}{\and}^2 \hat{a}_j^2,
\end{equation}
where $j$ labels the lattice sites and the bosonic operators $\hat{a}_j^\dag$ ($\hat{a}_j$) create (annihilate) an atom localised on site $j$. The hopping between neighbouring lattice sites is quantified by $J>0$, and $U>0$ is the pairwise on-site interaction strength. We consider a ring-shaped lattice of size $M$, such that $\hat{a}_{M+1} = \hat{a}_1$. Impurity atoms of a different species are then embedded in the Bose-Hubbard lattice \cite{Klein05}, with at most one impurity atom per lattice site. Setups introducing multiple atomic impurities into a large Bose gas have been demonstrated experimentally \cite{Catani12,Spethmann12,Catani09}. Only the two lowest internal states of each impurity are assumed to contribute to the dynamics, so that each impurity can be treated as a qubit. The excited and ground states are labelled $|1\ra$ and $|0\ra$ respectively. If the qubits are coupled to the Bose gas via a contact density-density interaction then the total Hamiltonian has the form \cite{Elliott16,Streif16}
\begin{equation}
	\label{eqn:total Ham}
	\hat{H} = \hat{H}_{BH} + \frac{\omega_0}{2}\sum_{j=1}^d \hat{\sigma}^{z}_j + \eta\sum_{j=1}^d |1\ra\la 1|_j \otimes \hat{n}_{l_j},
\end{equation}
where $d$ is the number of impurity atoms.

The first and second terms are the free Hamiltonians of the Bose gas \eqref{eqn:BH Hamiltonian} and the qubits respectively. The third term describes qubit-gas interactions, with the excited state of the qubit on site $l_j$ coupling to the bosonic number operator $\hat{n}_{l_j}=\hat{a}^\dag_{l_j} \hat{a}_{l_j}$ with strength $\eta\geq 0$. For simplicity, we assume that the qubits are degenerate and couple with equal strength to the gas, such that $\omega_0$ and $\eta$ are the same on all lattice sites.

The dynamics generated by the Bose-Hubbard Hamiltonian are generally complicated and approximations are needed in order to obtain analytic results. To this end, we assume that the gas is prepared in its ground state and work in the superfluid regime $U/J \ll 1$. In this case the Bogoliubov approximation \cite{Oosten01} can be applied to bring the Hamiltonian \eqref{eqn:total Ham} into the form
\begin{align}
	\label{eqn:ham bogo}
	&\hat{H} = \sum_k \omega_k \hat{b}_k^\dag \hat{b}_k + \frac{\overline{\omega}_0}{2}\sum_{j=1}^d \hat{\sigma}_j^{z} \\
	&+\eta  \sum_{j=1}^d \sum_k  \sqrt{\frac{\epsilon_kn_0}{\omega_k M}} |1\ra\la 1|_j\otimes \left(\hat{b}_k^\dag\ue^{ikal_j} + \hat{b}_k\ue^{-ikal_j}\right)\nonumber,
\end{align}
where we have neglected an overall constant term and introduced the renormalised transition frequency $\overline{\omega}_0 = \omega_0 + \eta n_0$ in terms of the quasicondensate density $n_0$. The operators $\hat{b}_k^\dag$ ($\hat{b}_k$) create (annihilate) Bogoliubov quasiparticles with quasimomentum $k$ and satisfy the canonical commutation relations. The sums are taken over $k \neq 0$ and the quasiparticle dispersion relation $\omega_k = \sqrt{\epsilon_k^2 + 2Un_0 \epsilon_k}$ depends on the single-particle energies $\epsilon_k = 4J\sin^2\left(ka/2\right)$, where $a$ is the lattice constant.

From here on we shall often refer to the ultracold gas as the `environment' and the subsystem of qubits as the `system'. We assume that the environment and the system are initially uncorrelated $\hat{\rho}(0) = \hat{\rho}_S(0)\otimes \hat{\rho}_E$, where $\hat{\rho}_S(0)$ is an arbitrary $d$-qubit state and $\hat{\rho}_E$ is the ground state of the Bose-Hubbard Hamiltonian \eqref{eqn:BH Hamiltonian}. With the Bogoliubov approximation it is then possible to derive an analytic expression for system state $\hat{\rho}_S(t) = \textnormal{tr}_E[\ue^{-it\hat{H}}\hat{\rho}(0)\ue^{it\hat{H}}]$, where $\hat H$ is the Hamiltonian \eqref{eqn:ham bogo} and the partial trace is taken over the environment. The matrix elements of the system state in the eigenbasis of the free qubit Hamiltonian are found to be of the form
\begin{equation}
\label{eqn:approx rho}
    [\hat{\rho}_S(t)]_{ij} = \ue^{-\gamma_{ij}(t)}\ue^{i\varphi_{ij}(t)}[\hat{\rho}_S(0)]_{ij},
\end{equation}
where $i = (i_1,\ldots,i_d)$ labels a basis vector with $i_j \in \{0,1\}$, and the time-dependent parameters $\gamma_{ij}(t)$ and $\varphi_{ij}(t)$ are real valued. A derivation of this result is provided in Appendix \ref{appendix:approxsol} but there are several details worth noting here.

First, the time evolution of $\hat{\rho}_S$ consists of a unitary and a non-unitary part. In the unitary part, we find that the coupling of the qubits to the ultracold gas induces a time-dependent qubit-qubit interaction with terms of the form $c_{ij}(t)\hat{\sigma}_i^z\hat{\sigma}_j^z$. That is, each qubit couples to all the others through a $ZZ$ interaction with a time-dependent strength that varies between different pairs of qubits. We shall see that this interaction can generate a nonlocal state from a separable initial state. To give some insight into how this can happen, consider an interaction of the form $g\hat{\sigma}_1^z\hat{\sigma}_2^z$ with $g\in\R$. Two qubits initialised in the spin-$x$ up state $|+,+\ra$ are mapped to the state $\cos(gt)|+,+\ra -i \sin(gt)|-,-\ra$. When $gt = \pi/4$ the separable initial state has evolved into a maximally entangled state.
\begin{figure*}[ht]
    \centering
		\includegraphics[width=0.98\textwidth]{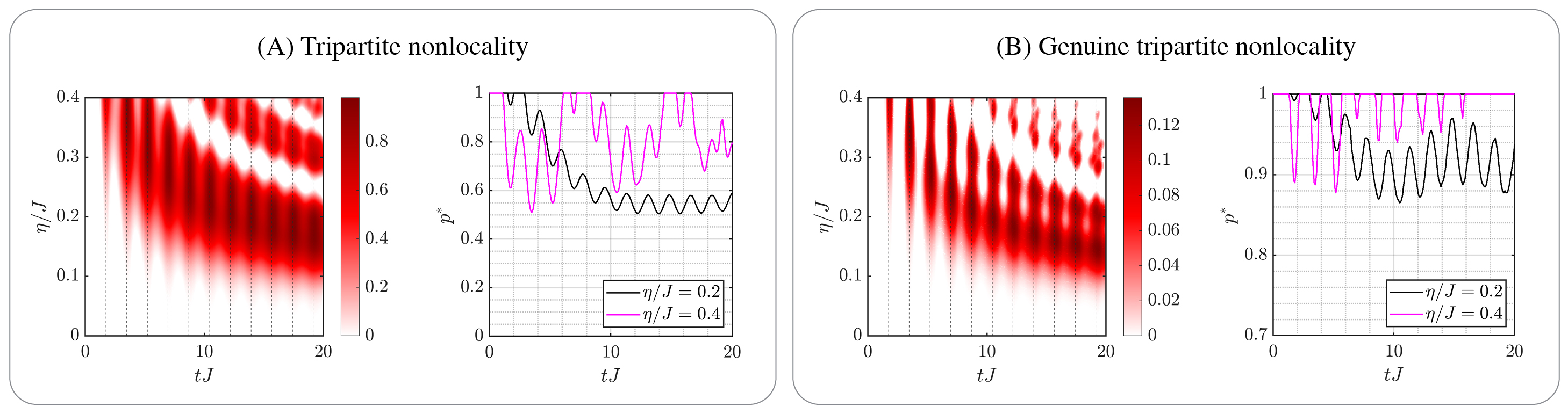}
	\caption{Generation of tripartite nonlocality (A) and genuine tripartite nonlocality (B) between three qubits coupled to a three-site Bose-Hubbard ring. Bell inequality violation vs.\ time and coupling strength is shown on the left of each panel. Positive values indicate violation. Noise robustness is shown on the right as $p^\ast$ vs.\ time. Smaller $p^\ast$ indicates greater robustness. The results are for an interacting ultracold gas of $N=100$ particles and interaction strength $UN=2J$. The qubit energy is $\omega_0=1$.}
	\label{fig:3m3i}
\end{figure*}

On the other hand, the non-unitary part gives rise to dephasing described by the parameters
\begin{equation}
\label{eqn:gamma param}
    \gamma_{ij}(t) = \sum_k \nu_k\sin^2\left(\frac{\omega_k t}{2}\right)\left|\sum_{m=1}^d \ue^{ikal_m}(i_m-j_m)\right|^2,
\end{equation}
with $\nu_k = 2\eta^2 n_0 \epsilon_k / \omega_k^3M$. This only affects the off-diagonal elements of the state and generally degrades any nonlocal correlations that may have built up. However, due to the sinusoidal factor, these parameters need not be monotonically increasing. There can be times when the time rate of change $\dot{\gamma}_{ij}(t)<0$, corresponding to a backflow of information from the environment to the system.

To understand this, recall that the trace distance can be viewed as a measure of distinguishability between two quantum states. If the trace distance, and hence distinguishablity, between two states is increasing at some instant of time, then there must be a flow of information from the environment to the system \cite{Breuer09}. In Appendix \ref{appendix:nonmarkov} we show that if $\dot{\gamma}_{ij}(t)<0$ at some time $t$ for some pair of indices $(i,j)$, then one can always find a pair of initial states such that the trace distance distance between them is increasing at time $t$. 

The parameters $\gamma_{ij}(t)$ are proportional to $\eta^2$ and, as expected, the strength of the dephasing is amplified when the interaction between the system and environment is increased. Furthermore, the coefficients $c_{ij}(t)$ in the $ZZ$ interaction are also proportional to $\eta^2$. We therefore expect there to be a trade-off between generating nonlocality slowly over an extended period of time, and quickly generating nonlocality that only lasts for a short period of time, due to a large dephasing rate. 

Finally, from the solution \eqref{eqn:approx rho} we see that the eigenstates of $\hat{\sigma}_1^z\dots\hat{\sigma}_d^z$ are stationary states of the approximate dynamics. Thus, to observe nonlocality generation, and non-trivial time evolution, the initial state cannot be a product of spin-$z$ eigenstates. Guided by the remarks above equation \eqref{eqn:gamma param} we shall assume that all the qubits are initially prepared in the spin-$x$ up state $|+,\dots,+\ra$.

In general, to determine whether a state is nonlocal is a challenging problem in the multipartite setting \cite{Brunner2014}. All tight correlation Bell inequalities for $d$ parties, each with a choice of two dichotomic measurements, were derived by Werner and Wolf \cite{Werner01} and \.{Z}ukowski and Brukner \cite{Zukowski02}. These $2^{2^d}$ inequalities can be expressed as a single nonlinear inequality
\begin{equation}
\label{eqn:WWZB}
    \sum_{r}|\tilde{\xi}(r)| \leq 1
\end{equation}
which we refer to as the WWZB inequality. Here $\tilde{\xi}(r)=2^{-d}\sum_{s}(-1)^{r\cdot s}\xi(s)$, $r,s$ are vectors in $\{0,1\}^d$ and $\xi(s)=\la \hat{A}_{s_1}^{(1)}\dots\hat{A}_{s_d}^{(d)}\ra$, where the components $s_j$ label the choice of the $\pm 1$-valued observable $\hat{A}_{s_j}^{(j)}$ made on subsystem $j$.

Various forms of nonlocality can be distinguished in the multipartite setting. For instance, it is possible for nonlocal correlations to be local with respect to some bipartition. On the other hand, when all parties are nonlocally correlated the correlations are said to be genuine multipartite nonlocal (GMNL). This is the strongest form of multipartite nonlocality. However, it is generally difficult to test if a state is GMNL \cite{Svetlichny87,Brunner2014}. Here we make use of the Bell inequality introduced in \cite{Bancal13}, a violation of which detects genuine tripartite nonlocality (GTNL), and find that GTNL can be established between three impurity atoms. For brevity we refer to this inequality as the GTNL inequality and provide the definition in Appendix \ref{appendix:gtnl}. 

Let us first examine a three-site Bose-Hubbard ring with a single impurity atom coupled to each site. In this case, we can numerically solve the Schr\"{o}dinger equation with the many-body Hamiltonian \eqref{eqn:total Ham} exactly to obtain the reduced state of the three qubits. We test the WWZB inequality \eqref{eqn:WWZB} using projective spin measurements $\hat{A}_{s_j}^{(j)}= \mathbf{u}_{s_j}\cdot \hat{\boldsymbol{\sigma}}_j$ on each of the $j\in\{1,2,3\}$ qubits. Here $\hat{\boldsymbol{\sigma}}_j$ is a vector of Pauli operators acting on qubit $j$ and the unit vector $\mathbf{u}_{s_j}$ is the Bloch vector of the measurement. We also test the GTNL inequality with projective spin measurements (see Appendix \ref{appendix:gtnl}). At each instant of time a maximisation of each inequality was performed over the twelve angles parameterising the measurements.

In panels A and B of Fig.~\ref{fig:3m3i} we plot respectively the degree of violation of the WWZB and GTNL inequalities against coupling strength and time. We see that nonlocal states can be generated from separable states across a wide range of parameters. Furthermore, across the entire parameter space we see oscillations in the degree of Bell inequality violation, which is a non-Markovian effect due to information backflow from the environment. Using \eqref{eqn:gamma param} we find that $\dot{\gamma}_{ij}(t)<0$ if and only if $\sin(\omega_k t)<0$, where $\omega_k$ is evaluated at $k=2\pi/3a$. Information thus flows back and forth between the system and environment, and information flow from the environment to the system vanishes at the times $t = 2n\pi/\omega_k$ with $n\in \N$. These are indicated by dashed-black vertical lines and coincide with the revivals of nonlocality.

For coupling strength $\eta \approx 0.2J$, we observe that nonlocality is generated and persists for a long period of time. This illustrates the trade-off between dephasing and unitary dynamics. Due to the non-Markovianity of the dynamics, the information lost to the environment quickly returns and complete loss of nonlocality is avoided.

The noise robustness of nonlocality can be quantified by subjecting $\hat{\rho}_S(t)$ to completely depolarising noise \cite{Almeida07}
\begin{equation}
    \label{eqn:robust def}
    \tilde{\rho}(p) = p\hat{\rho}_S(t) + (1-p)\frac{\hat{I}}{2^d}.
\end{equation}
%The largest value $p^\ast$ such that $\tilde{\rho}(p)$ is local for any $p\leq p^\ast$ provides a measure of robustness.
The smallest value $p^\ast$ such that $\tilde{\rho}(p^\ast)$ is nonlocal provides a measure of robustness. Panel A in Fig.~\ref{fig:3m3i} illustrates that robust nonlocality ($p^\ast \approx 0.5$) can be generated quickly for larger $\eta$ but is short lived. For weaker coupling strength, nonlocality builds up more slowly but remains robust over a longer time. In frame B of Fig. \ref{fig:3m3i} we see that the GTNL is less robust against noise but the behaviour with respect to $\eta$ is similar to the tripartite nonlocal case. 

\begin{figure}[t]
	\begin{center}
		\includegraphics[width=0.49\textwidth]{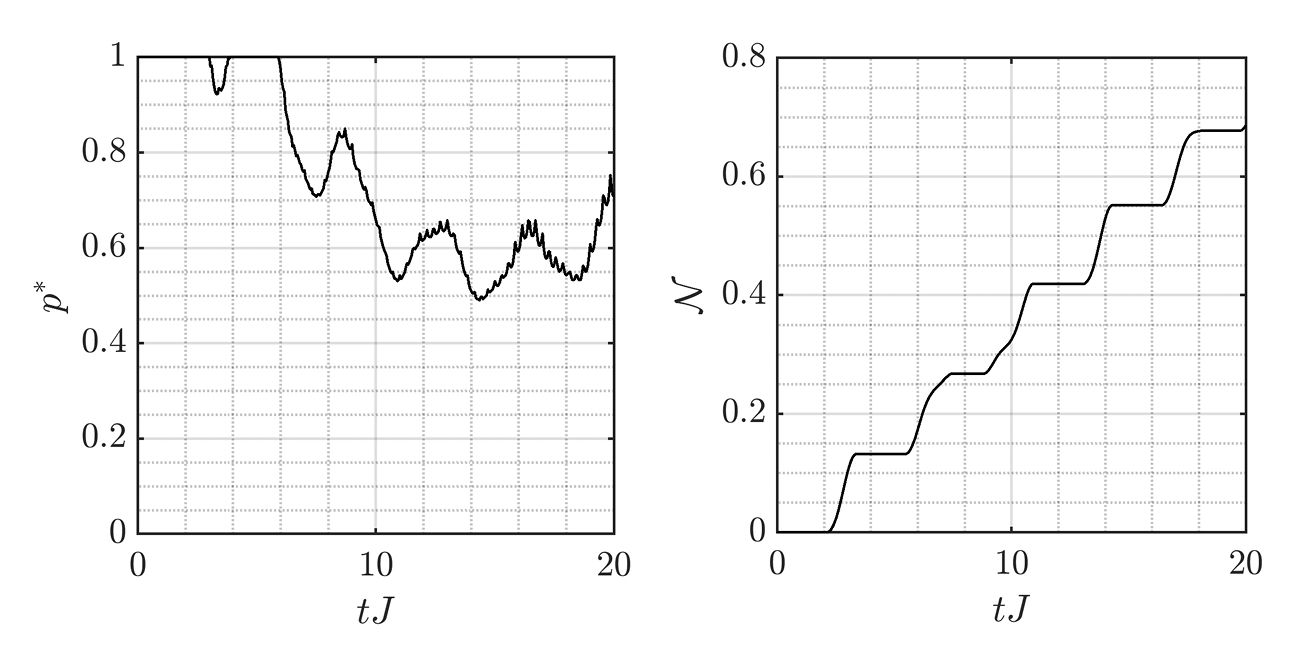}
	\end{center}
	\caption{Generation of nonlocality among five qubits coupled to a five-site Bose-Hubbard ring with strength $\eta=0.05J$. (Left) The noise robustness of nonlocality vs.\ time. Smaller $p^\ast$ indicates greater robustness. (Right) Non-Markovianity vs.\ time. Results are for a gas of $N=1000$ particles with interaction strength $UN=2J$ and qubit energy $\omega_0=1$.}
	\label{fig:5qubits}
\end{figure}

Using the approximate solution \eqref{eqn:approx rho}, we are also able to investigate nonlocality generation beyond the regime where exact numerical solutions are possible. The approximate solution is expected to be valid provided that the ultracold gas remains close to the ground state. We compared the approximate and exact numerical solutions for small, numerically tractable systems and found excellent agreement between the two. In Fig.~\ref{fig:5qubits} we consider a five-site Bose-Hubbard ring with one qubit coupled to each site. We again observe that separable states can evolve into robust nonlocal states that violate the WWZB inequality for projective spin measurements.

To illustrate the importance of non-Markovianity on the nonlocality generation we also plot the Breuer-Laine-Piilo (BLP) measure of non-Markovianity $\mathcal{N}$ \cite{Breuer09} in the right panel of Fig.~\ref{fig:5qubits}. This quantifies the total amount of information flowing back into the system from the environment and can be written as
\begin{equation}
    \mathcal{N} = \max_{\hat{\rho}_{1,2}(0)} \sum_i \left[D(\hat{\rho}_1(b_i),\hat{\rho}_2(b_i))-D(\hat{\rho}_1(a_i),\hat{\rho}_2(a_i))\right].
\end{equation}
 The sum is taken over all time intervals $(a_i,b_i)$ where the trace distance $D(\hat{\rho}_1(t),\hat{\rho}_2(t))$ between the two initial states states $\hat{\rho}_{1,2}(0)$ is increasing. Thus, whenever $\mathcal{N}$ increases there is information flowing from the environment to the system. In Fig. \ref{fig:5qubits} we observe that the oscillations in the robustness parameter $p^\ast$ coincide with the information flow into and out of the system. This confirms the intuition that memory effects are crucial in preventing the complete loss of nonlocality.
\begin{figure}[t]
	\begin{center}
		\includegraphics[width=0.49\textwidth]{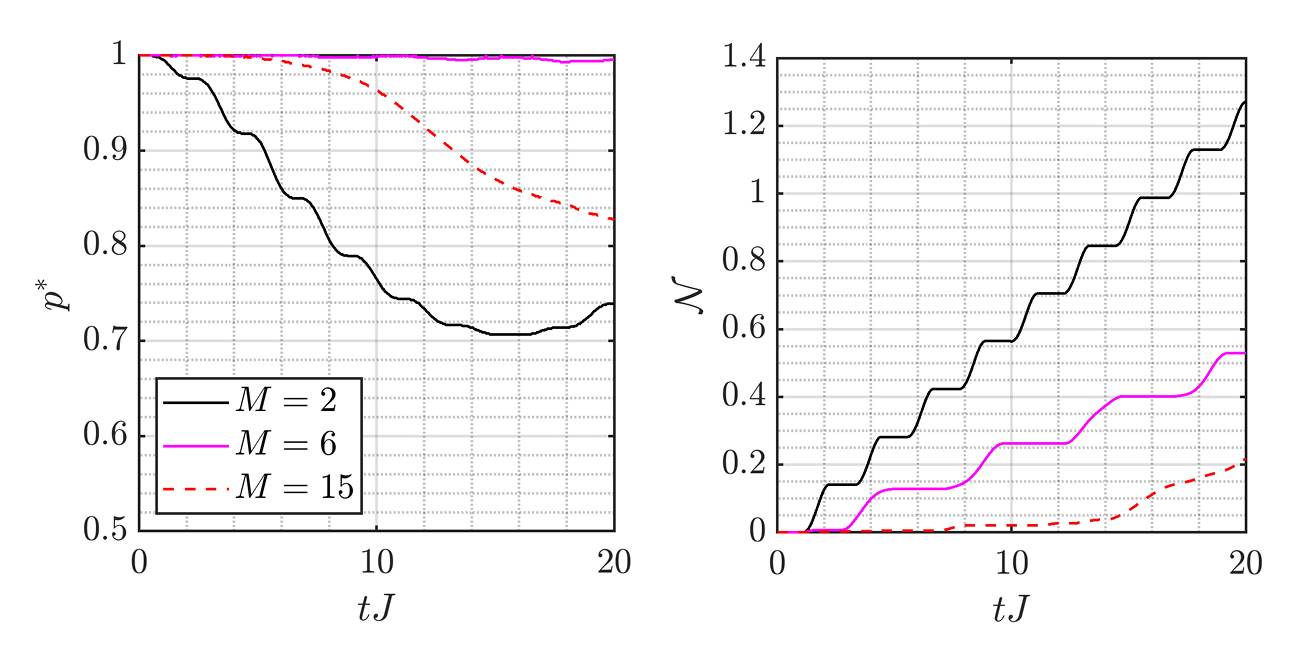}
	\end{center}
	\caption{Two impurity atoms embedded in a lattice of size $M$ at sites $1$ and $M$ respectively. (Left) The noise robustness of nonlocality vs.\ time when the non-unitary part of the dynamics is neglected. Smaller $p^\ast$ indicates greater robustness. (Right) Non-Markovianity vs.\ time. Results are for a gas of $N=1000$ particles with interaction strength $UN=2J$, coupling strength $\eta=0.04J$ and qubit energy $\omega_0=1$.}
	\label{fig:mm2i}
\end{figure}

The examples above focused on the case $d=M$, with one impurity embedded in each lattice site. When the number of lattice sites is increased, with the number of impurities and other parameters fixed, generating nonlocal correlations in a short time can become difficult. There are two main reasons for this. The first concerns the unitary part of the dynamics, generated by the effective Hamiltonian (see Appendix \ref{appendix:approxsol})
\begin{equation}
\label{eqn:effective h}
    \hat{h}(t) = \sum_{j=1}^d \omega_j(t) \hat{\sigma}_j^z + \sum_{j>m}^d \sum_{m=1}^{d-1}c_{jm}(t)\hat{\sigma}_j^z\hat{\sigma}_m^z.
\end{equation}
This has a complicated dependence on the system parameters, as illustrated in the left panel of Fig. \ref{fig:mm2i} for two impurities embedded on sites $1$ and $M$ respectively. Within the time scale considered, we observe that the unitary part generates increasingly weaker nonlocality as the lattice size increases from $M=2$ to $M=6$, at which point essentially zero nonlocality is generated. Nonlocal correlations appear again when $M$ is increased beyond this point. Furthermore, the dynamics of the impurities become `more Markovian' (with a smaller BLP measure of non-Markovianity) when the lattice size increases \cite{Sarkar20}. This is illustrated in the right panel of Fig. \ref{fig:mm2i}. A larger $M$ results in longer time intervals of information loss to the environment and less information backflow. Fig. \ref{fig:mm2i} shows how these two effects can combine together to suppress nonlocality generation at short times. However, at longer times strong persistent nonlocal correlations can be generated, even when the separation between the impurities is large. This raises the question of whether persistent nonlocality is possible in large lattices.

There is no condensate in an infinite one-dimensional lattice. Taking the continuum limit with the substitutions $\sum_k \to M\int_{-1/2}^{1/2}dq$ and $k \to 2\pi q/a$ leads to a divergent momentum integral in the expression for the condensate density \cite{Oosten01}. In a large but finite system, this divergence can be avoided by introducing a low-momentum cut off $q_0 = 1/M$, where the number of lattice sites $M$ is large. This yields the dephasing parameters
\begin{equation}
\label{eqn:gamma param cont}
   \gamma_{ij}(t) = 4\eta^2n_0\int_{q_0}^{1/2} \frac{\epsilon_q}{\omega_q^3}\sin^2\left(\frac{\omega_q t}{2}\right)S_{ij}(q)\,dq.
\end{equation}
Here $\epsilon_q = 4J\sin^2(\pi q)$, $\omega_q = \sqrt{\epsilon_q^2+2Un_0\epsilon_q}$ and
\begin{equation}
S_{ij}(q) = \left|\sum_{m=1}^d \ue^{2i\pi ql_m}(i_m-j_m)\right|^2.
\end{equation}
The same substitutions can also be used to obtain a large lattice expression for the term $\varphi_{ij}(t)$ in the approximate solution \eqref{eqn:approx rho}.

For simplicity we focus on the dynamics of two impurity atoms embedded in a large lattice. In this case we can apply the Horodecki criterion to quantify the degree of Bell inequality violation \cite{Horodecki95,Miranowicz04} (see Appendix \ref{appendix:horodecki}), avoiding the need to optimise over measurements. For a pair of qubits there are three different decoherence parameters that appear in the density matrix. The first, $\Gamma_0(t) \equiv \gamma_{(1,1),(1,0)}(t)$, is given by
\begin{equation}
    \Gamma_0(t)  = 4\eta^2n_0\int_{q_0}^{1/2}\frac{\epsilon_q}{\omega_q^3}\sin^2\left(\frac{\omega_q t}{2}\right) dq
\end{equation}
and is the dephasing rate that a single impurity atom experiences. The other two, $\Gamma_\pm(t)$, which are defined as $\gamma_{(1,1),(0,0)}(t)$ and $\gamma_{(1,0),(0,1)}(t)$ respectively, have the form $\Gamma_\pm(t) = 2\Gamma_0(t) \pm \Gamma(t)$ with
\begin{equation}
    \Gamma(t) = 8\eta^2n_0\int_{q_0}^{1/2}\frac{\epsilon_q}{\omega_q^3}\sin^2\left(\frac{\omega_q t}{2}\right)\cos\left(2\pi q(l_1-l_2)\right)\,dq.
\end{equation}
The decoherence rates $\Gamma_\pm(t)$ are not simply twice the single impurity dephasing rate. Rather, one is larger and one is smaller than $\Gamma_0(t)$. The system is said to exhibit super and subdecoherence. This feature of collective decoherence is known to occur when two qubits interact with a shared bosonic environment via a spin-boson type interaction \cite{Palma96,Cirone09}. 

\begin{figure}[t]
	\begin{center}
		\includegraphics[width=0.49\textwidth]{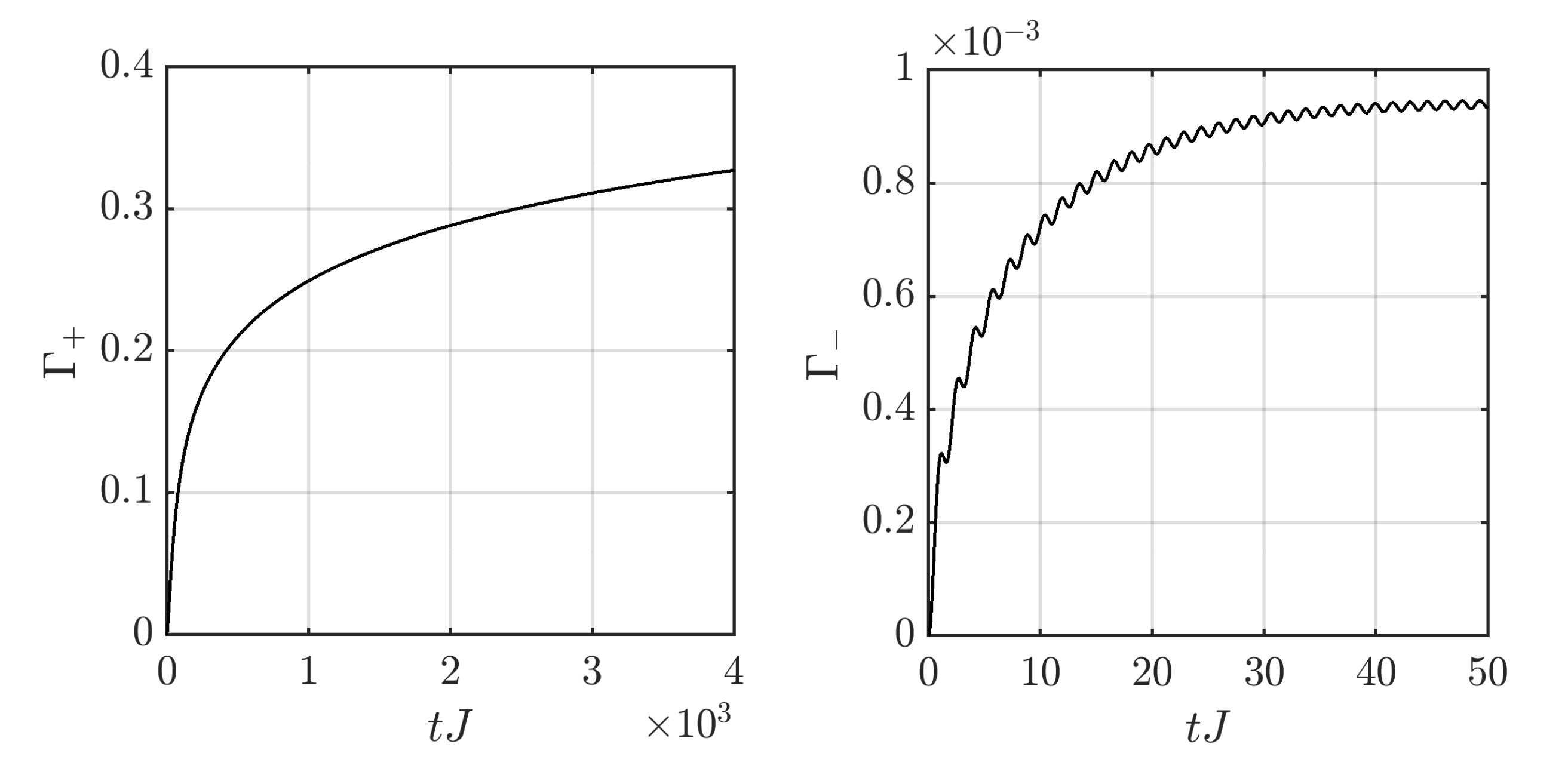}
	\end{center}
	\caption{Decoherence rates $\Gamma_+$ (left) and $\Gamma_-$ (right) vs.\ time. Results show the long-time evolution for a gas with $n=1$ atoms per lattice site, interaction strength $U=0.04J$, coupling strength $\eta=0.03J$ and qubit energy $\omega_0=1$. The low-momentum cut off $q_0 = 10^{-6}$, corresponding to $10^6$ lattice sites.}
	\label{fig:gam plus minus}
\end{figure}

\begin{figure}[t]
	\begin{center}
		\includegraphics[width=0.49\textwidth]{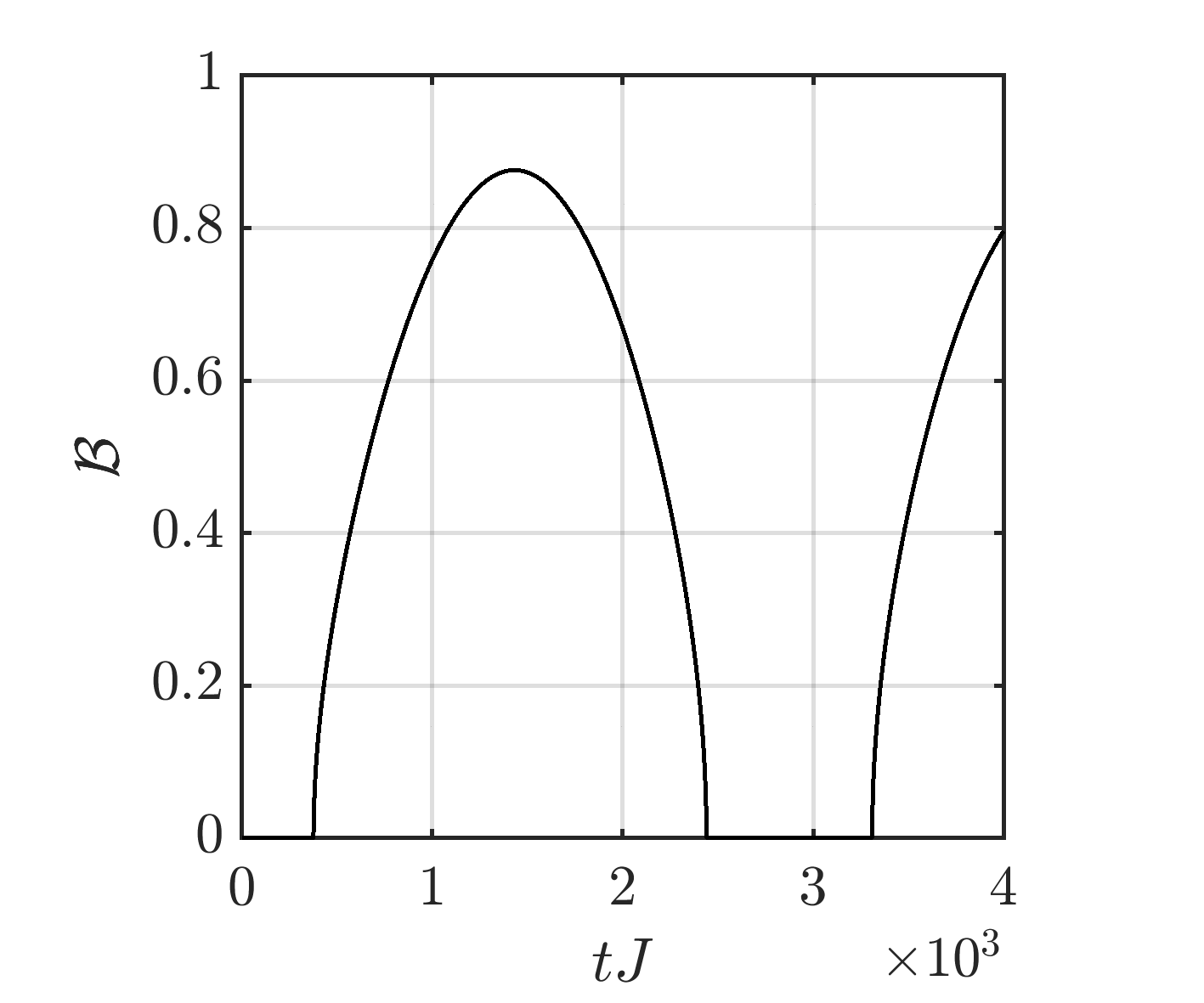}
	\end{center}
	\caption{Nonlocality generation in a large system. Measure of Bell inequality violation $\mathcal{B}$ vs.\ time. Values greater than zero indicate Bell inequality violation, and $\mathcal{B} = 1$ corresponds to a maximal violation of the CHSH inequality for some set of measurements. The parameters are the same as Fig. \ref{fig:gam plus minus}}
	\label{fig:nonlocal therm}
\end{figure}

The time evolution of $\Gamma_\pm(t)$ is plotted in Fig. \ref{fig:gam plus minus} for a pair of qubits embedded in neighbouring lattice sites. The rate $\Gamma_-(t)$ exhibits small oscillations, indicating (weak) non-Markovian behaviour. More importantly, the rates level out around a finite value, resulting in the persistence of coherences in the impurity subsystem at long times. A similar effect has previously been observed in a set up consisting of impurity atoms, each trapped in a double-well potential and interacting with a homogeneous Bose gas \cite{Cirone09}. Because the coherences are preserved, long-time generation and persistence of nonlocality is possible. This is illustrated in Fig. \ref{fig:nonlocal therm} for two impurities coupled to a large lattice with $10^6$ sites.


In conclusion, we have demonstrated that the system-environment coupling of impurity qubits in a cold Bose gas can generate multipartite, nonlocal correlations that persist over long times. The setup can be realised with current technology and the procedure is simple in the sense that no dynamic driving or control is needed, except for initialisation and read out. We have shown that robust 3- and 5-partite nonlocality can be observed (tolerating up to $\sim 50\%$ depolarising noise) as well as genuinely tripartite nonlocality. We note that the observation of (genuine) multipartite nonlocality also witnesses the presence of (genuine) multipartite entanglement. The number of parties studied here is limited by numerics, but we expect that nonlocality between an arbitrary number of parties is possible.

To gain insight into the many-body dynamics we made the Bogoliubov approximation and derived an approximate solution for the reduced state of the impurities. The time evolution was found to consist of two parts: an effective unitary dynamics, generating the nonlocality; and a non-unitary part, resulting in dephasing that tends to destroy the nonlocal correlations. We observed that non-Markovian effects in the reduced-state dynamics are crucial for maintaining nonlocality over long periods of time. Here, memory effects suppress the complete loss of information to the environment. This feature could be particularly useful in experimental setups that require flexibility in the manipulation of the impurities. Furthermore, we demonstrated that persistent nonlocality can still occur in large systems. This is possible due to the saturation of decoherence rates, resulting in sustained coherences in the impurity subsystem.

The analytic solution obtained using the Bogoliubov approximation is only expected to be valid provided that the ultracold gas remains close to its ground state. Surprisingly, in our numerical simulations we observed that the approximation can still give a reliable indication of nonlocality generation in situations where the trace distance between the exact numerical and approximate solutions is large. This suggests that the properties of the total state that the Bogoliubov approximation fails to account for may not play a crucial role in determining whether the reduced state is nonlocal. Exploring this connection could be beneficial for further understanding nonlocality generation in many-body systems.

Finally, there has been growing interest in the use of impurity atoms embedded in many-body environments, from nondestructive measurements of Bose-Einstein condensate phase fluctuations \cite{Bruderer06} and correlations in ultracold-atom systems \cite{Elliott16,Streif16}, to temperature measurements of Bose-Einstein condensates \cite{Hovhannisyan21}. In each of these examples information about a many-body system is extracted by making measurements on a probe system that typically has a simple structure, e.g., a two-level impurity atom. An interesting future research direction would be to examine whether nonlocal probe states can provide an advantage in applications of this kind.

We gratefully acknowledge support from the Carlsberg Foundation CF19-0313 and the Independent Research Fund Denmark 7027-00044B.

\bibliographystyle{quantum}
\bibliography{cold_nonlocal}

\appendix
\onecolumngrid

\section{Derivation of the approximate solution}
\label{appendix:approxsol}
Here we provide a derivation of the approximate solution \eqref{eqn:approx rho} in the main text. We refer to the subsystem of impurities as the `system' and the ultracold gas as the `environment'. We work in the zero-temperature limit and assume that the system and environment are initially uncorrelated $\hat{\rho}(0) = \hat{\rho}_S(0)\otimes \hat{\rho}_E$. Here $\hat{\rho}_S(0)$ is an arbitrary $d$-qubit state and $\hat{\rho}_E = |v\ra\la v|$ is the Bogoliubov vacuum state with $|v\ra = |0,\ldots,0\ra$.

The dynamics of the total system-environment state are generated by the Hamiltonian \eqref{eqn:ham bogo} in the main text, where the Bogoliubov approximation has been made. Setting
\begin{equation}
	\hat{H}_0 = \sum_k \omega_k \hat{b}_k^\dag \hat{b}_k + \frac{\overline{\omega}_0}{2}\sum_{j=1}^d \hat{\sigma}_j^z
\end{equation}
yields the interaction-picture Hamiltonian
\begin{equation}
	\hat{H}_I(t) = \sum_{j=1}^d\sum_k \xi_k |1\ra\la 1|_j \otimes \left(\hat{b}_k^\dag \ue^{i(\omega_k t + kal_j)}+h.c.\right),
\end{equation}
where we have defined $\xi_k = \eta\sqrt{n_0 \epsilon_k/\omega_k M}$. In order to solve the Schr\"odinger equation we use the Magnus expansion \cite{Magnus54} to obtain the time-evolution operator $\hat{U}_I(t)$. Fortunately, in this case the Magnus series truncates at the second term and we find $\hat{U}_I(t) = \exp(\hat{\Omega}_1(t)+\hat{\Omega}_2(t))$, where
\begin{equation}
	\label{eqn:omega1}
	\hat{\Omega}_1(t) = \sum_{j=1}^d \sum_k |1\ra\la 1|_j \otimes \left(\beta_{jk}(t)\hat{b}_k^\dag - h.c.\right),
\end{equation}
with $\beta_{jk}(t) = {\xi_k}\ue^{ikal_j}\left(1-\ue^{i\omega_k t}\right)/{\omega_k}$, and
\begin{align}
	\hat{\Omega}_2(t) = -i\sum_{j=1}^d &\left(\sum_{m=1}^d c_{jm}(t)\right)\hat{\sigma}_j^z \nonumber\\&-i\sum_{j>m}^d\sum_{m=1}^{d-1}c_{jm}(t)\hat{\sigma}_j^z\hat{\sigma}_m^z,
\end{align}
where we have neglected a term proportional to the identity and defined the time-dependent coefficients
\begin{equation}
	\label{eqn:c coeffs}
	c_{jm}(t) = \frac{1}{2}\sum_k\left(\frac{\xi_k}{\omega_k}\right)^2\left(\sin(\omega_k t)-\omega_k t\right)\cos(ka\left(l_j-l_m\right)).
\end{equation}

Note that $\hat{\Omega}_1$ and $\hat{\Omega}_2$ commute with each other. Furthermore, $\hat{\Omega}_2$ only acts on the system and also commutes with $\hat{H}_0$. Going back to the Schr\"odinger picture we therefore find that the coupling to the environment induces a unitary evolution $\hat{U}(t) = \exp(-i\hat{h}(t))$ on the subsystem of qubits with
\begin{equation}
	\label{eqn:eff unitary}
	\hat{h}(t) = \sum_{j=1}^d \omega_j(t) \hat{\sigma}_j^z + \sum_{j>m}^d \sum_{m=1}^{d-1}c_{jm}(t)\hat{\sigma}_j^z\hat{\sigma}_m^z,
\end{equation}
where $\omega_j(t) = \overline{\omega}_0t/2 + \sum_m c_{jm}(t)$.

To simplify the notation we define 
$\hat{\sigma}(t) = \hat{U}(t)\hat{\rho}_S(0)\hat{U}^\dag(t)$, which is the system state following the unitary evolution above. The total system-environment state at time $t$ can then be written as
\begin{equation}
	\hat{\rho}(t) = \hat{K}(t)\left(\hat{\sigma}(t)\otimes |v\ra\la v|\right)\hat{K}^\dag(t)
\end{equation}
where $\hat{K}(t) = \exp(-it\sum_k \omega_k \hat{b}_k^\dag \hat{b}_k)\exp(\hat{\Omega}_1(t))$. The system state is obtained by taking the partial trace over the environment
\begin{equation}
	\label{eqn:rhoS sum}
	\hat{\rho}_S(t) = \sum_{n=0}^\infty \la n| \ue^{\hat{\Omega}_1(t)}|v\ra \hat{\sigma}(t) \la v | \ue^{\hat{\Omega}_1^\dag(t)}|n\ra,
\end{equation}
where the sum is taken over all $n = (n_1,\ldots,n_{M-1})$. Because each of the different $k$ modes in $\hat{\Omega}_1(t)$ commute we can write
\begin{equation}
	\label{eqn:omega1 X}
	\ue^{\hat{\Omega}_1(t)} = \prod_k \exp\left(\hat{X}_k \otimes \hat{b}_k^\dag - \hat{X}_k^\dag \otimes \hat{b}_k\right)
\end{equation}
in terms of the operators $\hat{X}_k = \sum_{j=1}^d \beta_{jk}(t)|1\ra\la 1|_j$, where the $\beta_{jk}(t)$ are defined in \eqref{eqn:omega1}. The operators $\la n |\ue^{\hat{\Omega}_1(t)}|v\ra$ appearing in \eqref{eqn:rhoS sum} therefore factorise into products of operators of the form $\la n_k |\exp\left(\hat{X}_{k}\otimes \hat{b}_{k}^\dag - \hat{X}_{k}^\dag\otimes \hat{b}_{k}\right)|0\ra$, which, after applying the Baker-Campbell-Hausdorff formula, can be written as
\begin{equation}
	\frac{1}{\sqrt{n_k!}}\ue^{-\frac{1}{2}\hat{X}_k^\dag\hat{X}_k}\hat{X}_k^{n_k}.
\end{equation}
The $\hat{X}_k$ are diagonal in the standard basis, with time-dependent matrix elements
\begin{equation}
	x_{i}^k(t) = \frac{\xi_k}{\omega_k}\left(1-\ue^{i\omega_k t}\right)\sum_{m=1}^d i_m\ue^{ikal_m}.
\end{equation}
Here $i = (i_1,\ldots,i_d)$ with $i_j \in \{0,1\}$ is the matrix element index and $k$ labels the quasimomentum value. 

After a little algebra we obtain the matrix elements of the system state in the standard basis
\begin{equation}
	\label{eqn:rhoS glauber}
	[\hat{\rho}_S(t)]_{ij} = \prod_k \la x_j^k(t)|x_i^k(t)\ra [\hat{\sigma}(t)]_{ij},
\end{equation}
where $|x_j^k(t)\ra$ is a Glauber coherent state. It is straightforward to work out the action of the unitary part \eqref{eqn:eff unitary} on the system and obtain an expression for $[\hat{\sigma}(t)]_{ij}$. Inserting this into \eqref{eqn:rhoS glauber} finally gives
\begin{equation}
	\label{eqn:apndx rhos}
	[\hat{\rho}_S(t)]_{ij} = \ue^{-\gamma_{ij}(t)}\ue^{i\varphi_{ij}(t)}[\hat{\rho}_S(0)]_{ij}, 
\end{equation}
where the $\gamma_{ij}(t)$ are defined in equation \eqref{eqn:gamma param} of the main text and 
\begin{align}
	&\varphi_{ij}(t) = \sum_{s=1}^d \omega_s(t)\left((-1)^{i_s}-(-1)^{j_s}\right) \nonumber\\ &+\sum_{r>s}^d\sum_{s=1}^{d-1}c_{rs}(t)\left((-1)^{j_r+j_s}-(-1)^{i_r+i_s}\right)\nonumber\\
	&+\sum_k \sum_{r,s=1}^d f_k^2(t)\sin(ka(l_r-l_s))i_r j_s,
\end{align}
with $f_k(t) = 2\xi_k \sin\left(\omega_k t/2\right)/\omega_k$.

\section{Negative rates and non-Markovian dynamics}
\label{appendix:nonmarkov}
We show that if $\dot{\gamma}_{ij}(t)<0$ at some time $t$ for some pair of indices $(i,j)$, then the dynamics are non-Markovian. We shall do this by finding a pair of initial states such that their trace distance is increasing at time $t$.

First recall that the indices $i=(i_1,\ldots,i_d)$ and $j=(j_1,\ldots,j_d)$ label standard basis vectors, which we denote $|i\ra$ and $|j\ra$ respectively. Consider the pair of pure initial states
\begin{equation}
	|\psi_1(0)\ra = \frac{1}{\sqrt{2}}\left(|i\ra+|j\ra\right), \quad |\psi_2(0)\ra = \frac{1}{\sqrt{2}}\left(|i\ra-|j\ra\right),
\end{equation}
where the difference between the corresponding density operators is given by
\begin{equation}
	\hat{\Delta}(0) = \hat{\rho}_1(0) - \hat{\rho}_2(0) = |j\ra\la i| + |i\ra\la j|.
\end{equation}
We see that in the standard basis $\hat{\Delta}(0)$ is a matrix with $1$ in elements $(i,j)$ and $(j,i)$, and zeros elsewhere.

The time evolution of the difference is given by \eqref{eqn:apndx rhos} and is of the form
\begin{equation}
	\Delta(t) = \ue^{-\gamma_{ij}(t)}A(t).
\end{equation}
Here $A(t)$ is a Hermitian matrix with $\ue^{i\varphi_{ij}(t)}$ in element $(i,j)$, $\ue^{-i\varphi_{ij}(t)}$ in element $(j,i)$, and zeros elsewhere. The trace norm $\lVert A(t) \rVert_1 = \tr\left[\sqrt{A^\dag(t) A(t)}\right] = 2$. It then follows that the trace distance between the two states $\hat{\rho}_1(t)$ and $\hat{\rho}_2(t)$ is given by
\begin{equation}
	D(\hat{\rho}_1(t),\hat{\rho}_2(t)) = \frac{1}{2}\lVert \hat{\Delta}(t) \rVert_1 = \ue^{-\gamma_{ij}(t)}.
\end{equation}
Finally, we see that
\begin{equation}
	\frac{d}{dt}D(\hat{\rho}_1(t),\hat{\rho}_2(t)) = -\dot{\gamma}_{ij}(t)\ue^{-\gamma_{ij}(t)} > 0
\end{equation}
at time $t$ by hypothesis. The dynamics are therefore non-Markovian according to the BLP measure. 

\section{Genuine tripartite nonlocality}
\label{appendix:gtnl}
Here we provide the Bell-type inequality derived in \cite{Bancal13} to test for genuine tripartite nonlocality. Consider three parties, Alice, Bob and Charlie, each with a choice of two measurements labelled $x,y,z \in \{0,1\}$ with outputs $a,b,c \in \{0,1\}$. A violation of the inequality
\begin{align}
	I &= -2P(A_1 B_1) - 2P(B_1 C_1) - 2P(A_1 C_1)\nonumber\\ &- P(A_0 B_0 C_1) - P(A_0 B_1 C_0) - P(A_1 B_0 C_0) \nonumber\\ &+ 2P(A_1 B_1 C_0) + 2P(A_1 B_0 C_1) + 2P(A_0 B_1 C_1) \nonumber \\&+ 2P(A_1 B_1 C_1) \leq 0,
\end{align}
detects genuine tripartite nonlocality. Here, $P(A_i B_j) = P(a=0,b=0|x=i,y=j)$ and $P(A_i B_j C_k) = P(a=0,b=0,c=0|x=i,y=j,z=k)$.

The measurement operators defined in the main text have outcomes $\pm 1$, as required for testing the WWZB inequality. Measurement operators for projective spin measurements corresponding to output labels $a \in \{0,1\}$ take the form
\begin{equation}
	\hat{\Pi}_{a|x} = \frac{1}{2}\left[\hat I + (-1)^a \mathbf{u}_x \cdot \hat{\boldsymbol{\sigma}}\right],
\end{equation}
where the input $x\in\{0,1\}$ selects the unit vector $\mathbf{u}_x$ (the Bloch vector of the measurement). 

\section{The Horodecki criterion}
\label{appendix:horodecki}
In this appendix we provide the Horodecki criterion and a quantity to measure the degree of Bell inequality violation in a two-qubit system. The Horodecki criterion \cite{Horodecki95} provides a necessary and sufficient condition for a state $\hat{\rho}$ to violate the CHSH inequality \cite{Clauser69} for some set of measurements. Given a two-qubit state $\hat{\rho}$, construct the $3\times 3$ matrix $T$ with elements $T_{ij} = \tr[\hat{\sigma}_i\otimes\hat{\sigma}_j \hat{\rho}]$, where $\hat{\sigma}_i$ are the Pauli spin operators. Define $\mathcal{M}(\hat{\rho})$ as the sum of the two largest eigenvalues of $T^\dag T$. The Horodecki criterion states that $\hat{\rho}$ violates the CHSH inequality for some set of measurements iff $\mathcal{M}(\hat{\rho})>1$. To quantify the degree of Bell inequality violation we use the measure proposed in \cite{Miranowicz04}
\begin{equation}
	\mathcal{B}(\hat{\rho}) = \sqrt{\max\{0,\mathcal{M}(\hat{\rho})-1\}}.
\end{equation}
When $\mathcal{B}(\hat{\rho}) = 0$ there exists a local model and $\mathcal{B}(\hat{\rho}) = 1$ corresponds to a maximal violation of the CHSH inequality. The larger the value of $\mathcal{B}(\hat{\rho})>0$ the greater the violation of the CHSH inequality.

\end{document}